\documentclass[manuscript,screen,nonacm]{acmart}

\usepackage{blindtext}

\usepackage{xpatch}

\makeatletter
\xpatchcmd{\ps@firstpagestyle}{Manuscript submitted to ACM}{}{\typeout{First patch succeeded}}{\typeout{first patch failed}}
\xpatchcmd{\ps@standardpagestyle}{Manuscript submitted to ACM}{}{\typeout{Second patch succeeded}}{\typeout{Second patch failed}}    \@ACM@manuscriptfalse
\makeatother

\renewcommand\footnotetextcopyrightpermission[1]{} 
\setcopyright{none}
\usepackage{CJKutf8}
\usepackage{mathtools}
\usepackage{amsmath}
\usepackage{algorithm}
\usepackage{algorithmic}
\usepackage{graphicx}
\usepackage{amsfonts}
\usepackage{amsthm}
\usepackage{mathrsfs}
\usepackage{bbding}
\usepackage{ulem}
\usepackage{color}
\usepackage{bm}

\usepackage{ulem} 

\usepackage{hyperref}

\newtheorem{proposition}{Proposition}
\newtheorem{lemma}{Lemma}
\newtheorem{theorem}{Theorem}
\newtheorem{example}{Example}

\newcommand\Myperm[2][^n]{\prescript{#1\mkern-2.5mu}{}P_{#2}}

\definecolor{applegreen}{rgb}{0.55, 0.71, 0.0}

\newcommand{\MS}[1]{{{\color{black}#1}}}

\newcommand{\M}[1]{{{\color{black}#1}}}
\newcommand{\W}[1]{{{\color{black}#1}}}

\newcommand{\ms}[1]{{{\color{black}#1}}}

\usepackage{todonotes}

\AtBeginDocument{%
  \providecommand\BibTeX{{%
    \normalfont B\kern-0.5em{\scshape i\kern-0.25em b}\kern-0.8em\TeX}}}

\begin{document}

\title{A New Reduction Method from Multivariate Polynomials to Univariate Polynomials}


\author{Cancan Wang}
\email{wangcc@nbjl.nankai.edu.cn}
\author{Ming Su}
\authornotemark[1]
\email{nksuker@gmail.com}
\author{Gang Wang}
\email{wgzwp@163.com}
\affiliation{%
  \institution{College of Computer Science}
  \city{Tianjin}
  \country{China}
}

\author{Qingpo Zhang}
\affiliation{%
  \institution{Institute of Information Engineering, Chinese Academy of Sciences}
  \city{Beijing}
  \country{China}}
\email{zhangqingpo@126.com}

\begin{abstract}
  Polynomial multiplication is a fundamental problem in symbolic computation. There are efficient methods for the multiplication of two univariate polynomials. However, there is rarely efficiently nontrivial method for the multiplication of two multivariate polynomials. Therefore, we consider a new multiplication mechanism that involves a) reversibly reducing multivariate polynomials into univariate polynomials, b) calculating the product of the derived univariate polynomials by the Toom-Cook or FFT algorithm, and c) correctly recovering the product of multivariate polynomials from the product of two univariate polynomials. This work focuses on step a), expecting the degrees of the derived univariate polynomials to be as small as possible. We propose iterative Kronecker substitution, where smaller substitution exponents are selected instead of standard Kronecker substitution.
   We also apply the Chinese remainder theorem to polynomial reduction and find its advantages in some cases. Afterwards, we provide a hybrid reduction combining the advantages of both reduction methods.
 Moreover, we compare these reduction methods in terms of  lower and upper bounds of the degree of the product of two derived univariate polynomials, and their computational complexities.
  With randomly generated multivariate polynomials, experiments show that the degree of the product of two univariate polynomials derived from the hybrid reduction can be reduced even to approximately $3\%$ that resulting from the standard Kronecker substitution, implying an efficient subsequent multiplication of two univariate polynomials.
\end{abstract}

%



\maketitle

\section{Introduction} \label{sec_introduction}

The computation of polynomials
is important in many applications, such as factorization~\cite{1,2},  finding common factors and finding roots ~\cite{3}, and polynomial multiplication is a crucial problem. Various techniques for the fast multiplication of univariate polynomials have been studied, e.g., the Toom-Cook~\cite{4} method with O($\ell^{1+ \epsilon }$) \ms{arithmetic} complexity and classic FFT~\cite{5,6,7,8} series algorithms with
\ms{the bit complexity}
of $O(\ell \log{\ell} \log{ \log {\ell}} \log{p})$, where $ 0< \epsilon <1$, $p$ is a prime,   the degrees of \ms{two univariate polynomials over $\mathbb{F}_p$ are less than $\ell$}.
 D. Harvey,  J. Hoeven, and  G. Lecerf ~\cite{9} presented a new bound \W{O($\ell \log{\ell} \ 8^{\log^{*} {\ell}} \log{p} $)}, where $\log^{*} {\ell}=\min \left\{i \in N: \underbrace{ \log{\cdots \log}}_{i}{\ell}\leq 1 \right\} $.\ Then, D. Harvey and
   J. Hoeven ~\cite{10} improved this complexity to \W{O($\ell \log{\ell} \ 4^{\log^{*} {\ell}} \log{p} $)} bit operations. \ms{\cite{STOC2012} and \cite{STOC2016} discussed the nearly-optimal sparse fourier transform, and V. Nakos proposed a nearly optimal algorithm for the sparse univariate polynomial multiplication \cite{IT2020}.}

Fast computation over a polynomial ring with multivariates is fundamental in symbolic computation, and there are many useful applications. For example, solving systems of multivariate polynomial equations has been proven to be NP-complete. Accordingly, those schemes based on multivariate polynomials are considered good candidates for post-quantum cryptography. As another example,
\ms{simplifying algebraic systems over finite fields by using Gr{\"{o}}bner basis},
in which the multiplication of multivariate polynomials is a basic operation.
There is rarely effective method for the fast multiplication of multivariate polynomials, so we study a new mechanism for the multiplication of multivariate polynomials, which can be combined with existing optimized libraries on the multiplication of univariate polynomials, such as FFTW~\cite{11,12,13,14}.
Our multiplication mechanism consists of a) polynomial reduction: reversibly reducing multivariate polynomials into univariate polynomials, b) calculating the product of the derived univariate polynomials, and c) recovering the product of multivariate polynomials from the derived univariate product. The running time of step b) depends on the degree \W{of the product of} the derived univariate polynomials; thus, this paper will aim at step a), effective polynomial reduction methods that make the \W{degrees} of the univariate polynomials as small as possible, where the smaller the degrees of the derived univariate polynomials are, the faster the multiplication of the univariate polynomials, and the more effective the reduction method.

Regarding reduction from multivariate polynomials into univariate polynomials, Kronecker~\cite{15} proposed \textit{standard Kronecker substitution}, which replaces variable $x_i$ with $x^{\mathbb{D}^{i-1}}$ ($i \geq 1$), where $\mathbb{D}$ is a sufficient large constant. Later, Arnold Sch{\"{o}}nhage~\cite{16} extended Kronecker substitution to reduce the multiplication in $Z[x]$ to the multiplication in $Z$.
\ms{\cite{17} determined the substitution exponents based on the degrees of variables in the multivariate polynomials, while our proposed iterative substitution method is more dynamic,  whose substitution exponents selected for each variable are not only related to their degrees, but also related to monomials contained in the polynomials after substitution in the recent round. Therefore, the degree of the univariate polynomial corresponding to the product of two multivariate polynomials after substitution is always not larger than that in \cite{17}, and they are equal in some cases.}

In 2009, Harvey~\cite{18} proposed a new variant, multipoint Kronecker substitution, which selects $r$ ($r=2 \ \textrm{or} \ 4$) evaluation points, resulting in $r$ multiplications in $Z$, with each approximately $1/r$-th the scale of the original integer multiplication. In 2014, Arnold and Roche~\cite{19} proposed randomized Kronecker substitution, which randomly selects $v$ tuples of substitution exponents but does not address the reduction collisions.
Additionally, there are some other tips to process polynomials, e.g., extracting common factors and reducing the sparsity of the original polynomials~\cite{20}.

The major contributions of this paper are as follows:
\begin{itemize}
\item[1)] Based on standard Kronecker substitution, we propose \textit{iterative Kronecker substitution}, choosing smaller substitution exponents to minimize the degree of the derived univariate polynomial and running time of the subsequent multiplication. Additionally, we estimate the bounds of the degree (Theorem \ref{theorem_KS_degree}) by iterative Kronecker substitution and
    \ms{the optimal order of substitution should follow a {\it straight-pattern}, i. e.,  variables included in the multivariate polynomial should be reduced to the same variable}
    (Theorem \ref{theorem_straight_pattern}).
\item[2)] We apply the \textit{Chinese remainder theorem} (\textit{CRT}) to the polynomial reduction problem, which results in a smaller degree \W{of the product} of the derived univariate polynomials in some cases.
\item[3)] We predict the degree \W{of the product} of the univariate polynomials derived from iterative Kronecker substitution and CRT reduction and then propose a \textit{hybrid reduction} combining the advantages of methods 1) and 2).
\end{itemize}

The article is organized as follows. In Section \ref{sec_preliminaries}, we give some notations and introduce standard Kronecker substitution. In Section \ref{sec_proposed}, we propose three reduction algorithms: iterative Kronecker substitution, CRT reduction and hybrid reduction. At the end of Section \ref{sec_proposed}, we also compare four reduction methods on the lower and upper bounds of the degree of the product of two univariate polynomials and their computational complexity,
 and experimental results showing the efficiency of our proposed reduction methods are provided later in Appendix. Finally, we conclude in Section \ref{sec_conclusions}.

\section{Preliminaries}\label{sec_preliminaries}
\subsection{Notations}
Suppose the number of variables is $n$, and $f(x_1, \dots , x_n)$ and $g(x_1, \dots , x_n)$ are multivariate polynomials \ms{over finite fields}; then, $h(x_1, \dots , x_n)=f(x_1, \dots , x_n)g(x_1, \dots , x_n)$ is the $n$-variable product. We denote the univariate polynomials derived from $f(x_1, \dots , x_n)$ and $g(x_1, \dots , x_n)$ by $f(x)$ and $g(x)$, respectively, and the univariate product by $h(x)=f(x)g(x)$. In addition, we denote the degree of a polynomial $F(x_1, \dots , x_n)$ on variable $x_i$ by $\deg_{x_i}F$  ($1 \leq i \leq n$). Additionally, we denote the degrees of the original $f(x_1, \dots , x_n)$, $g(x_1, \dots ,$  $ x_n)$ and $h(x_1, \dots , x_n)$ on $x_i$  by $d_{f_{x_i}}$, $d_{g_{x_i}}$ and $d_{h_{x_i}}$ and the degrees of \W{univariate polynomials} $f(x)$, $g(x)$ and $h(x)$ by $d_{f_x}$, $d_{g_x}$ and $d_{h_x}$, respectively. It is straightforward that
\begin{equation} \label{eqn_dhx}
d_{h_{x_i}}=d_{f_{x_i}}+d_{g_{x_i}}, \quad d_{h_x}=d_{f_x}+d_{g_x}.
\end{equation}
Specifically, we denote the degree of the univariate product derived from standard Kronecker substitution by $d_{h_x}^{\textrm{SKS}}$, the degree from iterative Kronecker substitution by $d_{h_x}^{\textrm{IKS}}$, the degree from CRT reduction by $d_{h_x}^{\textrm{CRT}}$, and the degree from hybrid reduction by $d_{h_x}^{\textrm{HR}}$. If a monomial $ax_1^{k_1}\cdots x_n^{k_n}$ ($a \neq 0$) appears in polynomial $F$, we say that $x_1^{k_1}\cdots x_n^{k_n} \in F$ or $x_1^{k_1}\cdots x_n^{k_n}$ is \textit{contained} in $F$. We define
\[\mathit{Max}_F(i,j)=\max_{\substack{ \forall \left(x_i^{k_i}x_j^{k_j}\cdots \right)  \in  F }} { \left\{ k_j-k_i \right\} }.\]

\subsection{Standard Kronecker Substitution}

From~\cite{15,18}, the most important factor of standard Kronecker substitution is the large constant
\begin{equation} \label{eqn_D}
\mathbb{D}=\max_i {\left\{ d_{h_{x_i}} \right\}+1}.
\end{equation}
Let ${\mathbb{D}}_i={\mathbb{D}}^{i-1}$. The substitutions replace $x_i$ with $x^{{\mathbb{D}}_i}$  and yield the univariate polynomials
\begin{equation*}\label{eqn_sks}
f(x)= f\left(x^{{\mathbb{D}}_1}, \dots ,x^{{\mathbb{D}}_{n}}\right) \quad \textrm{and} \quad g(x) = g\left(x^{{\mathbb{D}}_1},\dots ,x^{{\mathbb{D}}_{n}}\right)
\end{equation*}
from $n$-variable $f(x_1, \dots , x_n)$ and $g(x_1, \dots , x_n)$. For monomials $t_0=x_1^{k_1^{(0)}} \cdots x_n^{k_n^{(0)}}$ and $t_1=x_1^{k_1^{(1)}}\cdots x_n^{k_n^{(1)}}$ contained in $f(x_1, \dots , x_n)$ or $g(x_1, \dots , x_n)$, the exponents of $x$ after the substitutions are
\begin{eqnarray} \label{Eqn-SKS-Substition}
K_{SKS}^{(0)} & = & k_1^{(0)}{\mathbb{D}}_1+\cdots +k_n^{(0)}{\mathbb{D}}_{n}  \quad \textrm{and} \quad K_{SKS}^{(1)}=k_1^{(1)}{\mathbb{D}}_1+\cdots +k_n^{(1)}{\mathbb{D}}_{n},
\end{eqnarray}
the ${\mathbb{D}}$-digit representations of which are exactly $\left( k_1^{(0)}k_2^{(0)} \cdots k_n^{(0)} \right)$ and $\left( k_1^{(1)}k_2^{(1)} \cdots k_n^{(1)} \right)$. Additionally, if
\begin{equation*}\label{eqn_sksK}
\MS{k_j^{(0)}=k_j^{(1)}\quad \textrm{for} \quad i+1\leq j \leq n, \quad}
\textrm{and} \quad k_i^{(m)}>k_{i}^{(1-m)},
\end{equation*}
then $K_{SKS}^{(m)}>K_{SKS}^{(1-m)}$ ($m \in \{0, 1\}$).  This fact implies no reduction collisions, i.e., different $n$-variable monomials contained in $f(x_1, \dots , x_n)$, $g(x_1, \dots , x_n)$ or $h(x_1, \dots , x_n)$ are not reduced into the same univariate monomial. Conversely, the inverse of standard Kronecker substitution transforms every univariate monomial $t=x^K$ into an $n$-variable monomial
\[t=x_1^{\left(K \bmod {\mathbb{D}} \right)}x_2^{\left(\lfloor K/{\mathbb{D}}\rfloor \bmod {\mathbb{D}} \right)} \cdots x_n^{\left(\lfloor K/{\mathbb{D}}^{n-1} \rfloor \bmod {\mathbb{D}} \right)}.\]

\begin{proposition}\label{prop_sks}

\begin{eqnarray*} \label{Bound-SKS}
 d_{h_{x_n}}{\mathbb{D}}^{n-1} \leq d_{h_x}^{\textrm{SKS}} \ms{ \leq {\mathbb{D}}^{n}-1 }.
\end{eqnarray*}

\end{proposition}
\begin{proof}
There are monomials $x_1^{k_1}\cdots x_n^{d_{f_{x_n}}}$ contained in $f(x_1, \dots , x_n)$ and $x_1^{k'_1} \cdots x_n^{d_{g_{x_n}}}$ contained in  $g(x_1, \dots , x_n)$; then, we have $d_{h_x}^{\textrm{SKS}} =d_{f_x}+d_{g_x} \geq d_{h_{x_n}}{\mathbb{D}}^{n-1}$ because $x_n$ is replaced with
$x^{ {\mathbb{D}}^{n-1} }$.  \\
The degrees of $f(x)$ and $g(x)$ satisfy
\begin{eqnarray*}
& & d_{f_x} \leq d_{f_{x_1}}+d_{f_{x_2}}{\mathbb{D}}+\cdots +d_{f_{x_n}}{\mathbb{D}}^{n-1}, \\
\quad & & d_{g_x} \leq d_{g_{x_1}}+d_{g_{x_2}}{\mathbb{D}}+ \cdots +d_{g_{x_n}}{\mathbb{D}}^{n-1}.
\end{eqnarray*}
Therefore, from (\ref{eqn_dhx}) and (\ref{eqn_D}), we have
\begin{eqnarray*}
 \qquad \qquad \qquad \qquad \qquad d_{h_x} &\leq& d_{h_{x_1}}+d_{h_{x_2}}{\mathbb{D}} +\cdots +d_{h_{x_n}}{\mathbb{D}}^{n-1}  \\
 &\leq& ({\mathbb{D}}-1)+({\mathbb{D}}-1){\mathbb{D}}+\cdots +({\mathbb{D}}-1){\mathbb{D}}^{n-1} \ms{ \leq {\mathbb{D}}^{n}-1 }.
\end{eqnarray*}
\end{proof}


\section{Proposed Polynomial Reduction Methods}  \label{sec_proposed}
In step a) of our proposed mechanism, multivariate polynomials should be reversibly transformed into univariate polynomials. In this section, we propose iterative Kronecker substitution instead of the standard Kronecker substitution, choosing smaller substitution exponents. Then, we give the CRT reduction method, using the Chinese remainder theorem to reduce the multivariate polynomials. Finally, we give a special case of CRT reduction and thereby propose hybrid reduction, combining the advantages of both methods but not increasing the computational complexity. The computational complexity of step b) depends on the size of $d_{h_x}$ derived from a polynomial reduction method; the smaller the value of $d_{h_x}$ is, the more effective the reduction method.

\subsection{Iterative Kronecker Substitution}
To minimize $d_{h_x}$, we propose iterative Kronecker substitution instead of choosing a sufficiently large constant ${\mathbb{D}}$ at a time. In the $k$-th iteration, we replace an existing variable $x_{i_k}$ with another existing variable $x_{j_k}$ (we denote this substitution as $x_{i_{k}} \rightarrow x_{j_{k}}$). We denote the entire substitution sequence \ms{including $n-1$ rounds of substitutions} by
\[S: \ x_{i_1} \rightarrow x_{j_1}, \ x_{i_2} \rightarrow x_{j_2}, \ \dots, \ x_{i_{n-1}} \rightarrow x_{j_{n-1}}. \ ({i_k}, {j_k} \notin \{{i_1}, {i_2}, \dots . {i_{k-1}}\} \ \textrm{for  all }\ k \leq n-1)\]
For the substitution $x_{i_{k}} \rightarrow x_{j_{k}}$, define
\[\mathscr{D}_{i_1 \rightarrow j_1, i_2 \rightarrow j_2, \dots, i_k \rightarrow j_k}=\deg_{x_{j_k}}(f \cdot g)(x_{i_k},x_{i_{k+1}},\dots,x_{i_{n-1}},x_{j_{n-1}})+1,\]
\MS{where $x_{j_k} \in \{ x_{i_{k+1}},\dots,x_{i_{n-1}},x_{j_{n-1}} \}$.}
Then, we replace $x_{i_k}$ with $x_{j_k}^{\mathscr{D}_{i_1 \rightarrow j_1, i_2 \rightarrow j_2, \dots, i_k \rightarrow j_k}}$, reducing currently $(n-k+1)$-variable polynomials $f(x_{i_k},x_{i_{k+1}}, \dots, x_{i_{n-1}},x_{j_{n-1}})$ and $g(x_{i_k},x_{i_{k+1}},\dots,x_{i_{n-1}},x_{j_{n-1}})$ into $(n-k)$-variable polynomials $f(x_{j_k}^{\mathscr{D}_{i_1 \rightarrow j_1, i_2 \rightarrow j_2, \dots, i_k \rightarrow j_k}}, x_{i_{k+1}},\dots, x_{i_{n-1}},x_{j_{n-1}})$ and $g(x_{j_k}^{\mathscr{D}_{i_1 \rightarrow j_1, i_2 \rightarrow j_2, \dots, i_k \rightarrow j_k}},x_{i_{k+1}},\dots, \\ x_{i_{n-1}},x_{j_{n-1}})$.
The final univariate polynomials after \ms{$n-1$ rounds of}  substitutions are $f(x_{j_{n-1}})$ and $g(x_{j_{n-1}})$ with the variable $x_{j_{n-1}}$. The number of substitution sequences is $(\Myperm[n]{2}) * (\Myperm[n-1]{2})* \cdots *(\Myperm[2]{2}) =n!(n-1)!$,  \ms{where the partial permutation $\Myperm[n]{r}$ is the number of arrangements of $r$ items from $n$ objects}. Among them, we choose the \textit{optimal sequence} with the minimum $d_{h_x}=\deg_{x_{j_{n-1}}}h(x_{j_{n-1}})=\deg_{x_{j_{n-1}}}f(x_{j_{n-1}})+\deg_{x_{j_{n-1}}}g(x_{j_{n-1}})$. We divide all \MS{substitution} sequences into two categories: sequences with $x_{j_1}=x_{j_2}=\cdots =x_{j_{n-1}}$ are \textit{straight-pattern} sequences; all others are \textit{intermediate-pattern} sequences.
\MS{First, we discuss the bounds of the degree $d_{h_x}^{\textrm{IKS}}$.}

\begin{theorem}\label{theorem_KS_degree}
\ms{The degree $d_{h_x}^{\textrm{IKS}}$ satisfies:}
\begin{equation} \label{eqn_ite_KS}
\prod_{\substack{i=1}}^{n}d_{h_{x_i}} \leq d_{h_x}^{\textrm{IKS}} < \prod_{\substack{i=1}}^{n}(d_{h_{x_i}}+1).
\end{equation}
\end{theorem}
\begin{proof}
\MS{Denote $\mathscr{R}_m^k$ by a set of variables having a path connected to the variable $x_m$ within the $k$ rounds of substitution sequence $x_{i_1} \rightarrow x_{j_1}, \ \dots, \ x_{i_{k}} \rightarrow x_{j_{k}}$.}
To prove Theorem \ref{theorem_KS_degree}, we first prove that after the first $k$ substitutions, for any remaining variable $x_m \in \{x_{i_{k+1}}, \dots, x_{i_{n-1}}, x_{j_{n-1}}\}$ ($x_{j_{k}} \in \{x_{i_{k+1}}, \dots, x_{i_{n-1}}, x_{j_{n-1}}\}$), the current degree of $(f \cdot g)$ on $x_m$ satisfies
\begin{equation}\label{eqn_toprove}
\prod_{\substack{s \in \{ x_m\} \cup \mathscr{R}_m^k }}d_{h_{s}} \leq \deg_{x_m}(f \cdot g)(x_{i_{k+1}}, \dots, x_{i_{n-1}}, x_{j_{n-1}}) < \prod_{\substack{ s \in \{ x_m\} \cup \mathscr{R}_m^k }}(d_{h_{s}}+1),
\end{equation}
by mathematical induction.
\begin{itemize}
\item Suppose $k=0$, which means that no substitution occurs. For any variable $x_m \in \{x_{i_1},\dots, x_{i_{n-1}}, x_{j_{n-1}}\}=\{x_1,x_2,\dots ,x_n\}$, $\mathscr{R}_m^0=\varnothing$, so (\ref{eqn_toprove}) holds.
\item Supposing (\ref{eqn_toprove}) holds for $k>0$ implies that
\begin{equation}\label{eqn_k1}
\prod_{\substack{s \in \{ x_{i_{k+1}}\} \cup \mathscr{R}_{i_{k+1}}^k }}d_{h_{s}} \leq \deg_{x_{i_{k+1}}}(f \cdot g)(x_{i_{k+1}}, \dots, x_{i_{n-1}}, x_{j_{n-1}}) < \prod_{\substack{ s \in \{ x_{i_{k+1}}\} \cup \mathscr{R}_{i_{k+1}}^k }}(d_{h_{s}}+1),
\end{equation}
\begin{equation}\label{eqn_k2}
\prod_{\substack{s \in \{ x_{j_{k+1}}\} \cup \mathscr{R}_{j_{k+1}}^k }}d_{h_{s}} \leq \deg_{x_{j_{k+1}}}(f \cdot g)(x_{i_{k+1}}, \dots, x_{i_{n-1}}, x_{j_{n-1}}) < \prod_{\substack{ s \in \{ x_{j_{k+1}}\} \cup \mathscr{R}_{j_{k+1}}^k }}(d_{h_{s}}+1).
\end{equation}
During the $(k+1)$-th substitution, the substitution exponent
\begin{equation}\label{eqn_k3}
\mathscr{D}_{i_1 \rightarrow j_1,\dots,i_{k+1} \rightarrow j_{k+1}} = \deg_{x_{j_{k+1}}}(f \cdot g)(x_{i_{k+1}},\dots, x_{i_{n-1}}, x_{j_{n-1}})+1.
\end{equation}
In addition,
\begin{eqnarray*}
\deg_{x_{j_{k+1}}}(f \cdot g)(x_{j_{k+1}}^{\mathscr{D}_{i_1 \rightarrow j_1,\dots,i_{k+1} \rightarrow j_{k+1}}},  x_{i_{k+2}}, \dots, x_{i_{n-1}}, x_{j_{n-1}})= \max \Big\{a \times \mathscr{D}_{i_1 \rightarrow j_1,\dots,i_{k+1} \rightarrow j_{k+1}} \\ +b \Big| \   (x_{i_{k+1}}^{a} x_{j_{k+1}}^{b} \MS{\cdots} )  \in (f \cdot g)(x_{i_{k+1}},\dots, x_{i_{n-1}}, x_{j_{n-1}})\Big\}.
\end{eqnarray*}
Thus,
\begin{eqnarray*}
\mathscr{D}_{i_1 \rightarrow j_1,\dots,i_{k+1} \rightarrow j_{k+1}} \times \deg_{x_{i_{k+1}}}(f \cdot g)(x_{i_{k+1}}, \dots, x_{i_{n-1}}, x_{j_{n-1}}) \leq \deg_{x_{j_{k+1}}}(f \cdot g)(x_{j_{k+1}}^{\mathscr{D}_{i_1 \rightarrow j_1,\dots,i_{k+1} \rightarrow j_{k+1}}}, \nonumber \\   x_{i_{k+2}}, \dots, x_{i_{n-1}}, x_{j_{n-1}}) \leq \mathscr{D}_{i_1 \rightarrow j_1,\dots,i_{k+1} \rightarrow j_{k+1}} \times \deg_{x_{i_{k+1}}}(f \cdot g)(x_{i_{k+1}}, \dots, x_{i_{n-1}}, x_{j_{n-1}})+ \nonumber \\ \deg_{x_{j_{k+1}}}(f \cdot g)(x_{i_{k+1}}, \dots, x_{i_{n-1}}, x_{j_{n-1}}).
\end{eqnarray*}
Then, from (\ref{eqn_k1}), (\ref{eqn_k2}) and (\ref{eqn_k3}),  we have
\begin{eqnarray}\label{eqn_k4}
&&\Big( \!\!\!\!   \prod_{\substack{s \in \{ x_{j_{k+1}}\} \cup \mathscr{R}_{j_{k+1}}^k }} \!\!\!\!\!\!\!\! d_{h_{s}}+1 \Big) \times  \!\!\!\!\!\!\!\!\prod_{\substack{s \in \{ x_{i_{k+1}}\} \cup \mathscr{R}_{i_{k+1}}^k }} \!\!\!\!\!\!\!\!\!\!\!\!   d_{h_{s}} \leq \deg_{x_{j_{k+1}}}(f \cdot g)(x_{j_{k+1}}^{\mathscr{D}_{i_1 \rightarrow j_1,\dots, i_{k+1} \rightarrow j_{k+1}}}, x_{i_{k+2}}, \dots,  x_{i_{n-1}}, x_{j_{n-1}}) \nonumber \\
&& < \Big(\deg_{x_{j_{k+1}}}(f \cdot g)(x_{i_{k+1}}, \dots, x_{i_{n-1}}, x_{j_{n-1}}) +1 \Big) \times \Big(\deg_{x_{i_{k+1}}}(f \cdot g)(x_{i_{k+1}}, \dots, x_{i_{n-1}}, x_{j_{n-1}}) +1\Big)
\nonumber \\
&& \leq \prod_{\substack{s \in \{ x_{j_{k+1}}\} \cup \mathscr{R}_{j_{k+1}}^k }}(d_{h_{s}}+1) \times \prod_{\substack{s \in \{ x_{i_{k+1}}\} \cup \mathscr{R}_{i_{k+1}}^k }}(d_{h_{s}}+1).
\end{eqnarray}
With $\mathscr{R}_{j_{k+1}}^{k+1}=\mathscr{R}_{j_{k+1}}^{k} \cup \{x_{i_{k+1}}\} \cup \mathscr{R}_{i_{k+1}}^{k}$, \MS{$\mathscr{R}_{j_{k+1}}^{k} \cap \{x_{i_{k+1}}\} \cap \mathscr{R}_{i_{k+1}}^{k}=\varnothing$} and $j_{k+1} \in \{i_{k+2}, \cdots , i_{n-1},j_{n-1}\}$, (\ref{eqn_k4}) becomes
\begin{eqnarray*}
\prod_{\substack{s \in \{ x_{j_{k+1}}\} \cup \mathscr{R}_{j_{k+1}}^{k+1} }}d_{h_{s}} \leq \deg_{x_{j_{k+1}}}(f \cdot g)(x_{i_{k+2}}, \dots, x_{i_{n-1}}, x_{j_{n-1}}) < \prod_{\substack{ s \in \{ x_{j_{k+1}}\} \cup \mathscr{R}_{j_{k+1}}^{k+1} }}(d_{h_{s}}+1).
\end{eqnarray*}
For other variables $x_m \in \{x_{i_{k+2}},\dots, x_{i_{n-1}}, x_{j_{n-1}}\} \setminus \{x_{j_{k+1}}\}$, $\mathscr{R}_{m}^{k+1}=\mathscr{R}_{m}^{k}$ and \MS{the degree of $(f \cdot g)$} on $x_m$ remains. In conclusion, (\ref{eqn_toprove}) holds for $k+1$.
\end{itemize}
\MS{Finally, we obtain (\ref{eqn_ite_KS}) when $k=n-1$.}
\end{proof}

The following Theorem~\ref{theorem_straight_pattern} implies that the optimal sequence must follow a straight-pattern having $n!$ choices, reducing the search space from $n!(n-1)!$ to $n!$.
\begin{theorem} \label{theorem_straight_pattern}
The optimal substitution sequence must belong to the straight-pattern category.
\end{theorem}
\begin{proof} For any intermediate-pattern substitution sequence, there must be a straight-pattern substitution sequence with smaller (or the same) $d_{h_x}=\deg_{x_{j_{n-1}}}h(x_{j_{n-1}})$.
\begin{itemize}
\item[(1)] $n=3$. For any intermediate-pattern sequence
\[S_{intermediate}: x_{i_1} \rightarrow x_{j_1}, \ x_{j_1} \rightarrow x_{j_2}, \quad (i_2=j_1)\]
Let $\mathscr{D}_{i_1 \rightarrow j_1} = d_{h_{x_{j_1}}}+1$; we have
\begin{equation}\label{eqn_11}
\deg_{x_{j_1}}h(x_{j_1}^{\mathscr{D}_{i_1 \rightarrow j_1}}, x_{j_1}, x_{j_2})=\max\left\{k_{i_1} \times \mathscr{D}_{i_1 \rightarrow j_1}+k_{j_1}| (x_{i_1}^{k_{i_1}} x_{j_1}^{k_{j_1}} \cdot )  \in h(x_{i_1}, x_{j_1}, x_{j_2})\right\}.
\end{equation}
Let $\mathscr{D}_{i_1 \rightarrow j_1, j_1 \rightarrow j_2}= d_{h_{x_{j_2}}}+1$;  we have
\begin{eqnarray} \label{eqn_22}
&& \deg_{x_{j_2}}h(x_{j_2}^{\mathscr{D}_{i_1 \rightarrow j_1} \times \mathscr{D}_{i_1 \rightarrow j_1, j_1 \rightarrow j_2}}, x_{j_2}^{\mathscr{D}_{i_1 \rightarrow j_1, j_1 \rightarrow j_2}}, x_{j_2}) =   \nonumber \\
&& \max \!\!\Big\{ k_{i_1}\times \mathscr{D}_{i_1 \rightarrow j_1} \times \mathscr{D}_{i_1 \rightarrow j_1, j_1 \rightarrow j_2}+\!\! k_{j_1} \times \mathscr{D}_{i_1 \rightarrow j_1, j_1 \rightarrow j_2} +\!\! k_{j_2}  | x_{i_1}^{k_{i_1}} x_{j_1}^{k_{j_1}} x_{j_2}^{k_{j_2}} \in h(x_{i_1}, x_{j_1}, x_{j_2}) \Big\}.
\end{eqnarray}
Suppose there is a straight-pattern sequence
\[ S_{straight}: x_{j_1} \rightarrow x_{j_2}, \ x_{i_1} \rightarrow x_{j_2}.\]
Let $\mathscr{D}_{j_1 \rightarrow j_2} = d_{h_{x_{j_2}}}+1$; we have
\begin{equation}\label{eqn_33}
\deg_{x_{j_2}}h(x_{i_1}, x_{j_2}^{\mathscr{D}_{j_1 \rightarrow j_2}}, x_{j_2})=\max\left\{ k_{j_1} \times \mathscr{D}_{j_1 \rightarrow j_2} + k_{j_2}| (\cdot x_{j_1}^{k_{j_1}}x_{j_2}^{k_{j_2}}) \in h(x_{i_1},x_{j_1},x_{j_2})\right \}.
\end{equation}
Let $\mathscr{D}_{j_1 \rightarrow j_2, i_1 \rightarrow j_2} = \deg_{x_{j_2}}h(x_{i_1}, x_{j_2}^{\mathscr{D}_{j_1 \rightarrow j_2}}, x_{j_2})+1$; we have
\begin{eqnarray}\label{eqn_44}
    \deg_{x_{j_2}}h(x_{j_2}^{\mathscr{D}_{j_1 \rightarrow j_2, i_1 \rightarrow j_2}}, x_{j_2}^{\mathscr{D}_{j_1 \rightarrow j_2}}, x_{j_2}) &=& \max\Big\{k_{i_1} \times \mathscr{D}_{j_1 \rightarrow j_2, i_1 \rightarrow j_2} + k_{j_1} \times \mathscr{D}_{j_1 \rightarrow j_2} +k_{j_2} | x_{i_1}^{k_{i_1}}x_{j_1}^{k_{j_1}}x_{j_2}^{k_{j_2}} \nonumber \\
& &\in h(x_{i_1},x_{j_1},x_{j_2}) \Big\} .
\end{eqnarray}
In addition, (\ref{eqn_11}), (\ref{eqn_22}) and (\ref{eqn_44}) reach their respective maximum at the same monomial $t_1\!=\!x_{i_1}^{d_{h_{x_{i_1}}}} x_{j_1}^{K_{j_1}} x_{j_2}^{K_{j_2}}$, where $K_{j_1}=\max\left\{k_{j_1}| x_{i_1}^{d_{h_{x_{i_1}}}} x_{j_1}^{k_{j_1}} \in h(x_{i_1},x_{j_1},x_{j_2})\right\}$,  $K_{j_2}=\max\left\{k_{j_2}| x_{i_1}^{d_{h_{x_{i_1}}}} x_{j_1}^{K_{j_1}} x_{j_2}^{k_{j_2}} \in h(x_{i_1},x_{j_1},x_{j_2})\right\}$; (\ref{eqn_33}) reaches its maximum at another monomial $t_2=\cdot x_{j_1}^{d_{h_{x_{j_1}}}} x_{j_2}^{K'_{j_2}}$ ($K'_{j_2} \leq d_{h_{x_{j_2}}}$).
Then, we have
\begin{eqnarray*}
(\ref{eqn_22})-(\ref{eqn_44}) &=& \left(d_{h_{x_{i_1}}} \times (d_{h_{x_{j_1}}}+1) \times (d_{h_{x_{j_2}}}+1)+ K_{j_1} \times (d_{h_{x_{j_2}}}+1) + K_{j_2} \right) - \\
& &  \left(d_{h_{x_{i_1}}} \times (d_{h_{x_{j_1}}} \times (d_{h_{x_{j_2}}}+1)+K'_{j_2}+1) + K_{j_1} \times (d_{h_{x_{j_2}}}+1) + K_{j_2} \right) \\
&=& d_{h_{x_{i_1}}}d_{h_{x_{j_2}}}-d_{h_{x_{i_1}}}K'_{j_2}\geq 0,
\end{eqnarray*}
implying that $S_{straight}$ is better than $S_{intermediate}$.
\item[(2)] $n>3$.
Suppose $S_1$ is the optimal substitution sequence and belongs to the intermediate-pattern category. Then, $S_1$ would be
\[S_1: \underbrace{\dots}_{\textrm{part}_1}\ x_a \rightarrow x_b \underbrace{\dots}_{\substack{\textrm{part}_2: \  \textrm{substitutions}\\ \textrm{without} \ x_a \ \textrm{and} \ x_b}} x_b \rightarrow x_c \ \underbrace{\dots}_{\textrm{part}_3} \quad (a \neq b \neq c).\]
We can transform $S_1$ into
\[S_2: \underbrace{\dots}_{\textrm{part}_1}  \underbrace{\dots}_{\substack{\textrm{part}_2: \  \textrm{substitutions}\\ \textrm{ without } \ x_a \ \textrm{and} \ x_b}} { x_a \rightarrow x_b, x_b \rightarrow x_c } \ \underbrace{\dots}_{\textrm{part}_3}.\]
According to the above analysis with $3$ variables,
\[S_3: \underbrace{\dots}_{\textrm{part}_1}  \underbrace{\dots}_{\substack{\textrm{part}_2: \  \textrm{substitutions}\\ \textrm{ without } \ x_a \ \textrm{and} \ x_b}} { x_b \rightarrow x_c, x_a \rightarrow x_c }\ \underbrace{\dots}_{\textrm{part}_3}\]
is better than $S_2$.
\end{itemize}
This derives the contradiction, and the proof is complete.
\end{proof}

The search space for the optimal sequence is too large. Inspired by Theorem~\ref{theorem_straight_pattern}, we consider the following straight-pattern substitution sequence $S_{IKS}: x_2 \rightarrow x_1, \ x_3 \rightarrow x_1, \cdots , \ x_{n} \rightarrow x_1$.
\footnote{\MS{The ratio of $\frac {d_{h_x} \ \textrm{derived from} \ S_{IKS} } {d_{h_x} \textrm{derived from the optimal substitution sequence}}$ \ is in $[1, \ \frac{ \prod_{\substack{i=1}}^{n}(d_{h_{x_i}}+1)} {\prod_{\substack{i=1}}^{n}d_{h_{x_i}}})$ by Theorem \ref{theorem_KS_degree}, whose upper bound tends to be close to 1 provided that  $d_{h_{x_i}}$ is large enough for any $1\leq i \leq n$. }}
In addition, for simplicity, we let
\begin{equation}\label{eqn_iksd}
D_i=\mathscr{D}_{2 \rightarrow 1, 3 \rightarrow 1, \cdots , i \rightarrow 1}=\deg_{x_1}h \left(x_1, x_1^{D_2}, \dots , x_1^{D_{i-1}}, x_i, \dots, x_n\right)+1.
\end{equation}
Algorithm~\ref{alg_ite_KS}  shows the details of the iterative Kronecker substitution with substitution exponents from (\ref{eqn_iksd}). The inverse of the iterative Kronecker substitution is clarified in Proposition~\ref{lem_iiks}.
\begin{algorithm}
\footnotesize
\caption{Iterative-Kronecker-Substitution}
\label{alg_ite_KS}
\begin{algorithmic}[1]
\REQUIRE $n$-variable polynomials $f(x_1, \dots , x_n)$ and $g(x_1, \dots , x_n)$
\ENSURE univariate polynomials $f(x)$ and $g(x)$ and substitution exponents \{$D_i$\}
\STATE $D_1 \gets 1$
\FOR { $i=2 \to n$}
	\STATE $D_i \gets \deg_{x_1}f+\deg_{x_1}g+1$
	\FOR {each monomial $t=\left( x_1^{k_1}x_i^{k_i}x_{i+1}^{k_{i+1}} \cdots x_n^{k_n} \right)$ contained in  $f$ or $g$}
	\STATE $t \gets x_1^{\left(k_1+k_i \times D_i \right)}x_{i+1}^{k_{i+1}} \cdots x_n^{k_n}$
	\ENDFOR
\ENDFOR
\RETURN univariate polynomials $f(x_1)$, $g(x_1)$ and substitution exponents \{$D_i$\}
\end{algorithmic}
\end{algorithm}
\begin{proposition} \label{lem_iiks}
In the inverse of iterative Kronecker substitution, we can correctly recover the $(n-i+2)$-variable  product
\[h\left(x^{D_1}, \dots , x^{D_{i-1}}, \bm{x_{i}}, x_{i+1}, \dots , x_n\right)=(f\cdot g)\left(x^{D_1}, \dots , x^{D_{i-1}}, \bm{x_{i}},x_{i+1}, \dots , x_n\right)\]
from the $(n-i+1)$-variable product
\[h\left(x^{D_1}, \dots , x^{D_{i-1}}, x^{D_{i}},x_{i+1}, \dots , x_n\right)=(f\cdot g)\left(x^{D_1}, \dots , x^{D_{i-1}}, x^{D_{i}}, x_{i+1}, \dots , x_n\right) \]
by replacing every $(n-i+1)$-variable monomial $x^Kx_{i+1}^{k_{i+1}}\cdots x_n^{k_{n}}$ with the $(n-i+2)$-variable monomial
\[x^{(K \bmod D_i)}\bm{x_i^{\lfloor K/D_i \rfloor}}x_{i+1}^{k_{i+1}}\cdots x_n^{k_{n}}.\]
\end{proposition}


\begin{example} \label{Example_iks}
We provide one $3$-variable example to show the 3 steps of the proposed multiplication mechanism by iterative Kronecker substitution: $f(x_1, x_2,x_3)=x_1^7x_2^7x_3^7+x_1x_2^7x_3^{17}$ and $ g(x_1, x_2,x_3)=x_2^3x_3^{34}+x_1^{8}x_2^{8}x_3^8$.
\begin{itemize}
\item[step a)] \textbf{Polynomial Reduction.} From (\ref{eqn_iksd}), we have: \\
$
\begin{array}{lll}
D_2=16, & \ f(x, x^{16},x_3)=x^{119}x_3^7+x^{113}x_3^{17}, & \ \text{and} \quad g(x, x^{16},x_3)=x^{48}x_3^{34}+x^{136}x_3^8; \\
D_3=256, & \ f(x, x^{16},x^{256})=x^{1911}+x^{4465}, & \ \text{and} \quad g(x, x^{16},x^{256})=x^{8752}+x^{2184}.
\end{array}
$
\item[step b)] \textbf{Univariate Multiplication.} We calculate the univariate product $h(x)=f(x)g(x)=x^{10663}+x^{4095}+\bm{x^{13217}}+x^{6649}.$
\item[step c)] \textbf{Recovery.} By Proposition~\ref{lem_iiks}, we can recover \\
$
h(x,x^{16},x_3)=x^{167}x_3^{41}+ x^{255}x_3^{15}+x^{161}x_3^{51}+x^{249}x_3^{25}; \\
h(x_1,x_2,x_3) =x_1^{7}x_2^{10}x_3^{41}+ x_1^{15}x_2^{15}x_3^{15}+x_1x_2^{10}x_3^{51}+x_1^{9}x_2^{15}x_3^{25}=(f \cdot g)(x_1,x_2,x_3).
$
\end{itemize}
From (\ref{eqn_D}), the standard Kronecker substitution chooses ${\mathbb{D}}=52$ and $d_{h_x}^{\textrm{SKS}}=138425$, but $d_{h_x}^{\textrm{IKS}}=13217$.
\end{example}

\subsection{CRT Reduction}    \label{section_CRT}
We apply the Chinese remainder theorem to the polynomial reduction problem.
\begin{lemma}[Chinese remainder theorem] \label{lemma_CRT}
Let $p_1, p_2, \dots ,p_n$ be $n$ positive integers that are coprime in pairs, i.e., $gcd(p_i, p_j)=1$ for all $ i \neq j$. Let $M=p_1 \times p_2 \times  \cdots \times  p_n$, $(u_1, u_2, \dots , u_n)$ be an $n$-tuple satisfying $0 \leq u_i < p_i \ \textrm{for all} \ 1 \leq i \leq n$; then, there is a unique integer $U$, $0 \leq U < M$, that satisfies the following congruential relations:
\begin{eqnarray}\label{equation_1}
\left\{
\begin{array}{l}
U \equiv u_1 \pmod{p_1} \\
U \equiv u_2 \pmod{p_2} \\
\cdots \\
U \equiv u_n \pmod{p_n}
\end{array}
\right. .
\end{eqnarray}
Additionally, we have
\begin{equation}\label{eqn_CRTU}
U \equiv \left( \frac{M}{p_1}a_1 u_1 + \frac{M}{p_2}a_2 u_2 + \cdots +\frac{M}{p_n}a_n u_n \right) \ (\bmod \ M),
\end{equation}
where $a_i$ is the inverse of $\frac{M}{p_i}$, satisfying $\frac{M}{p_i}a_i \equiv 1 \ (\bmod \ {p_i})$ ($i=1,2, \cdots ,n$). See more details in \cite{21}.
\end{lemma}

It is clear that we have the following result.

\begin{lemma}[CRT addition]\label{lemma_CRTAdd}
Let $p_1, p_2, \dots ,p_n$ be $n$ positive integers that are coprime in pairs, $V \equiv v_i \pmod{ p_i}$ and $U \equiv u_i \pmod{ p_i}$; then,
\begin{equation*}
U+V \equiv u_i + v_i \pmod{ p_i}, \ \textrm{for all} \ 1 \leq i \leq n.
\end{equation*}
\end{lemma}

Assuming $\left\{p_i: \ p_i > d_{h_{x_i}},\ 1 \leq i \leq n \right\}$ are coprime in pairs, CRT reduction transforms every monomial $x_1^{u_1}x_2^{u_2} \cdots x_n^{u_n}$ contained in $f(x_1, \dots ,x_n)$ or $g(x_1, \dots , x_n)$ into the univariate $x^U$ by (\ref{eqn_CRTU}). Lemma~\ref{lemma_CRT}  and Lemma~\ref{lemma_CRTAdd} guarantee that two different $n$-variable monomials will not be reduced into the same univariate monomial, and the obtained $h(x)=f(x)g(x)$ is exactly the univariate polynomial reduced from $h(x_1, \dots ,x_n)$. Therefore, we can correctly recover $h(x_1, \dots , x_n)$ from the calculated $h(x)$ by (\ref{equation_1}).
	
As an example, we illustrate CRT reduction using polynomials from Example~\ref{Example_iks} \M{again}. First, we choose bases $p_1=17$, $p_2=31$ and $p_3=52$, which are coprime in pairs and satisfy $p_i>d_{h_{x_i}}$. From (\ref{eqn_CRTU}), the integers corresponding to tuples $(7,7,7)$, $(1,7,17)$, $(0,3,34)$ and $(8,8,8)$ are 7, 69, 34 and 8, respectively. Therefore, we have univariate $f(x)=x^{7} + x^{69} $ and $g(x)=x^{34} + x^{8}$ after CRT reduction. Then, we calculate $h(x)=f(x)g(x)= x^{41} + x^{15}+ \bm{x^{103}} + x^{77} $. By (\ref{equation_1}), we recover $h(x_1, x_2,x_3)=x_1^{(41 \bmod {17})}x_2^{(41 \bmod {31})}x_3^{(41 \bmod {52})}+\cdots=x_1^{7}x_2^{10}x_3^{41}+x_1^{15}x_2^{15}x_3^{15}+x_1x_2^{10}x_3^{51}+x_1^{9}x_2^{15}x_3^{25}$, which is exactly the product of $f(x_1, x_2,x_3)$ and  $g(x_1, x_2,x_3)$. The degree $d_{h_x}^{\textrm{CRT}}$ is 103, much smaller than
$d_{h_x}^{\textrm{IKS}}=13217$ and $d_{h_x}^{\textrm{SKS}}=138425$.

\begin{algorithm}[!htb]
\footnotesize
\caption{CRT-Reduction}
\label{alg_crtsub}
\begin{algorithmic}[1]
\REQUIRE $n$-variable polynomials $f(x_1, \dots , x_n)$ and $g(x_1, \dots , x_n)$
\ENSURE univariate polynomials $f(x)$, $g(x)$ and bases  $\{p_1, p_2, \dots ,p_n\}$
\FOR {$i=1 \to n$}
	\STATE $p_i \gets \deg_{x_i}f(x_1, \dots , x_n)+\deg_{x_i}g(x_1, \dots , x_n)+1$
\ENDFOR
\STATE 
\MS{Increasingly adjust} $p_1, p_2, \cdots ,p_n$ so that they are coprime in pairs
\STATE $ M \gets p_1 \times  p_2  \times   \cdots \times p_n $
\FOR {$i=1 \to n$}
	\STATE $ M_i \gets M/p_i $
	\STATE $a_i \gets M_i^{-1} \  \pmod {p_i}$
\ENDFOR
\FOR {each monomial $t=\left(x_1^{k_1}x_2^{k_2}\cdots x_n^{k_n} \right)$ contained in  $f(x_1, \dots , x_n)$ or $g(x_1, \dots , x_n)$}
	\STATE $K \gets \sum_{\substack{i=1}}^n M_i   a_i  k_i$
	\STATE $t \gets x^{\left(K \bmod M \right)}$
\ENDFOR
\RETURN univariate polynomials $f(x)$, $g(x)$ and bases  $\{p_1, p_2, \dots ,p_n\}$
\end{algorithmic}
\end{algorithm}

The CRT reduction and inverse of CRT reduction are explicitly described in Algorithm~\ref{alg_crtsub} and Algorithm~\ref{alg_inversecrtsub}. In Algorithm~\ref{alg_crtsub}, $M_i^{-1}$ is the inverse of $M_i=\frac{M}{p_i}$ satisfying $M_i^{-1} M_i \equiv 1 \ (\bmod \ p_i)$, calculated by the extended  Euclid's algorithm.

\begin{algorithm}[!htb]
\footnotesize
\caption{Inverse-CRT-Reduction}
\label{alg_inversecrtsub}
\begin{algorithmic}[1]
\REQUIRE univariate polynomial $h(x)$ and bases  $\{p_1, p_2, \dots ,p_n\}$
\ENSURE $n$-variable polynomial $h(x_1, \dots , x_n)$
\FOR {each monomial $t=x^K$ contained in  $h(x)$ }
	\STATE $t \gets x_1^{\left(K \bmod p_1 \right)}x_2^{\left(K \bmod p_2\right)} \cdots x_n^{\left(K \bmod p_n\right)}$
\ENDFOR
\RETURN $n$-variable polynomial $h(x_1, \dots , x_n)$
\end{algorithmic}
\end{algorithm}

Every $n$-variable monomial $x_1^{u_1}x_2^{u_2}\cdots x_n^{u_n}$ contained in  $f(x_1, \dots , x_n)$ or $g(x_1, \dots , x_n)$ is reduced to $x^{U}$ with $U$ in $[0, M)$; thus, $d_{h_x}^{\text{CRT}}$ is in $[0, 2M)$. The upper bound $2M$ is greater than $\prod_{\substack{i=1}}^n(d_{h_{x_i}}+1)$, the upper bound of $d_{h_x}^{\text{IKS}}$. \MS{Generally, CRT reduction is not as good as iterative Kronecker substitution in terms of the degree of the reduced univariate polynomial but is better in some typical cases, e.g., the case of $k_1=k_2=\cdots =k_n$ for every monomial $x_1^{k_1}\cdots x_n^{k_n}$ and the case of Example~\ref{Example_iks}.}

\subsection{Hybrid Reduction}   \label{section_combine}
In hybrid reduction, we consider CRT reduction as an alternative to iterative Kronecker substitution in every iteration. By predicting the degree of the univariate product derived from both methods, we select the appropriate one in every iteration. First, we consider the case of 2-variable polynomials to demonstrate how to estimate the size of the degree.

Let $f(x_1,x_2)$, $g(x_1,x_2)$ be two $2$-variable polynomials. According to (\ref{eqn_ite_KS}), we can use  $\prod_{\substack{i=1}}^2d_{h_{x_i}}$ to approximate $d_{h_x}^{\textrm{IKS}}$. Before estimating the size of $d_{h_x}^{\textrm{CRT}}$, we first introduce a special case of CRT reduction to reduce $2$-variable polynomials.
Define
\[p_1=\max{ \left\{ d_{h_{x_1}}+1, \ d_{h_{x_2}}+2 \right\} }, \quad p_2=p_1 - 1;\]
then, $p_1$ and $p_2$ are coprime and $p_i>d_{h_{x_i}}$, \MS{$i=1, 2$}. Let $x_1^{k_1}x_2^{k_2} $ be any monomial contained in $f(x_1,x_2)$ or $g(x_1,x_2)$. Then, $p_1 \times (k_2-k_1) +k_1 = (p_2+1) \times (k_2-k_1) +k_1 = p_2 \times (k_2-k_1)+k_2$.
According to (\ref{equation_1}), if $k_2-k_1 \geq 0$, the integer $K$ ($K<p_1 \times p_2$) corresponding to tuple $(k_1, k_2)$ is $\left(p_1 \times (k_2-k_1) +k_1 \right)$. Next, we attempt to remove this precondition ``if $k_2-k_1 \geq 0$". Multiply $f(x_1,x_2)$ by $x_2^{m_f}$ so that $(k_2+m_f)-k_1 \geq 0$ holds in every monomial  $x_1^{k_1}x_2^{k_2+m_f} $ contained in $x_2^{m_f} f(x_1,x_2)$, and similarly for $g(x_1,x_2)$ with $x_2^{m_g}$.
The minimal choices of $m_f$ and $m_g$ are
\begin{eqnarray*}
m_f = \mathit{Max}_f(2,1), \ \textrm{and} \ m_g = \mathit{Max}_g(2,1).
\end{eqnarray*}
Now, we change the CRT bases into $p'_1=\max{ \left\{ d_{h_{x_1}}+1, \ d_{h_{x_2}}+m_f+m_g+2 \right\} }$ and $p'_2=p'_1-1$. The integer corresponding to the tuple $(k_1, k_2+m_f)$ is
\begin{equation}\label{eqn_iks_K}
K_{CRT}=p'_1 \times (m_f+k_2-k_1) +k_1
\end{equation}
by (\ref{equation_1}). Reducing the modified $x_2^{m_f} f(x_1,x_2)$  and  $x_2^{m_g} g(x_1,x_2)$ into univariate $f'(x)$ and $g'(x)$ with bases $p'_1$ and $p'_2$, if ignoring the addition factor $k_1$ in (\ref{eqn_iks_K}), with $k_1$ being much smaller than $p'_1$, we have
\begin{eqnarray}
\label{eqn_f}\deg_xf'(x) &\approx & \left(\mathit{Max}_f(2,1) + \mathit{Max}_f(1,2) \right) \times p'_1  , \\
\label{eqn_g}\deg_xg'(x) &\approx & \left(\mathit{Max}_g(2,1) + \mathit{Max}_g(1,2) \right) \times p'_1.
\end{eqnarray}
We further estimate the degree of product $h'(x)=f'(x)g'(x)$  as the sum of (\ref{eqn_f}) and (\ref{eqn_g}).

\begin{algorithm}[!htb]
\footnotesize
\caption{Hybrid-Reduction}
\label{alg_combine}
\begin{algorithmic}[1]
\REQUIRE $n$-variable polynomials $f(x_1, \dots , x_n)$ and $g(x_1, \dots , x_n)$
\ENSURE univariate polynomials $f(x)$ and $g(x)$
\FOR {$r=2 \to n$}
	\STATE select two existing variables $x_i=x_1$ and $x_j=x_r$
	\STATE $d_i \gets \deg_{x_i}f+\deg_{x_i}g$
	\STATE $d_j \gets \deg_{x_j}f+\deg_{x_j}g$
	\STATE $m_f \gets \mathit{Max}_f(j,i) $, \quad  $m_g \gets \mathit{Max}_g(j,i)$
	\STATE $p_i \gets \max{\left\{d_i+1, d_j+2+m_f+m_g\right\}}$, \ $p_j \gets p_i -1$
	\STATE $d_{\textrm {CRT}} \gets \left(m_f + m_g + \mathit{Max}_f(i,j)  +  \mathit{Max}_g(i,j) \right) \times p_i$
	\IF{$d_{\textrm {CRT}} < d_i \times d_j$   }
    \item[]    $\rhd$ \textit{CRT branch}: select CRT reduction to reduce $x_i$ and $x_j$
	 \FOR {each monomial $t=\left(x_i^{k_i}x_j^{k_j}\cdots \right) $ contained in  $f$ }
			\STATE $t \gets x_i^{(m_f+k_j-k_i) \times p_i +k_i}\cdots  $
		\ENDFOR
		\FOR {each monomial $t=\left(x_i^{k_i}x_j^{k_j}\cdots \right) $ contained in  $g$ }
			\STATE $t \gets x_i^{(m_g+k_j-k_i) \times p_i +k_i}\cdots  $
		\ENDFOR
	\ELSE
      \item[] $\rhd${ \textit{iterative-Kronecker branch}: select iterative Kronecker substitution to reduce $x_i$ and $x_j$     }
		\STATE $D_j \gets d_i+1$
		\FOR {each monomial $t=\left(x_i^{k_i}x_j^{k_j}\cdots \right) $ contained in  $f$ or $g$ }
			\STATE $t \gets x_i^{\left(k_i+k_j \times D_j \right)}\cdots  $
		\ENDFOR
	\ENDIF
\ENDFOR
\RETURN univariate polynomials $f(x)$ and $g(x)$
\end{algorithmic}
\end{algorithm}

In each iteration of hybrid reduction, let the two involved variables be $x_i$ and $x_j$ ($i \neq j $), and {by (\ref{eqn_ite_KS}), (\ref{eqn_f}) and (\ref{eqn_g})}, we choose to use CRT reduction or iterative Kronecker substitution to reduce $x_i$ and $x_j$ depending on which one is smaller: $\deg_{x_i}(f \cdot g)(x_i,x_j, \dots)\times \deg_{x_j}(f \cdot g)(x_i,x_j, \dots)$ and
\begin{eqnarray}\label{eqn_dcrt}
\Big( \mathit{Max}_{f(x_i,x_j, \dots)}(j,i) + \mathit{Max}_{f(x_i,x_j, \dots)}(i,j) +\mathit{Max}_{g(x_i,x_j, \dots)}(j,i)  +  \mathit{Max}_{g(x_i,x_j, \dots)}(i,j) \Big)  \times \nonumber \\ \max{ \Big\{ \deg_{x_i}(f \cdot g)(x_i,x_j, \dots)+1, \ \  \deg_{x_j}(f \cdot g)(x_i,x_j, \dots)+\mathit{Max}_{f(x_i,x_j, \dots)}(j,i) +\mathit{Max}_{g(x_i,x_j, \dots)}(j,i) +2 \Big\}.}
\end{eqnarray}
Algorithm~\ref{alg_combine} shows the details of hybrid reduction, in which {line 8 to line 14} constitute the CRT branch, where CRT reduction is selected to reduce the two involved variables, and {line 15 to line 20} constitute the iterative-Kronecker branch, where iterative Kronecker substitution is chosen. In addition, the notation $d_{CRT}$ in line 7 in Algorithm~\ref{alg_combine} is exactly (\ref{eqn_dcrt}).

In the CRT branch of hybrid reduction, because $p_1=p_2+1$, we can avoid the inverse computing $\left(\frac{M}{p_i}\right)^{-1}  \pmod {p_i}$ and
calculate $K_{CRT}$ corresponding to $(k_1, k_2)$ in Lemma~\ref{lemma_CRT} by (\ref{eqn_iks_K}), rather than by (\ref{eqn_CRTU}). The running time of each iteration (both the CRT branch and iterative-Kronecker branch) is linear in the number of monomials contained in $f$ or $g$, similar to the \MS{standard} Kronecker substitution.

\subsection{Comparison}
Table~\ref{compare_dhx} shows the lower and upper bounds of the degree $d_{h_x}$ derived from standard Kronecker substitution, iterative Kronecker substitution, CRT reduction and hybrid reduction.
\begin{table}[H]
\footnotesize
\caption{Bounds of the degree $d_{h_x}$ derived from the four reduction algorithms}
\label{compare_dhx}
\tabcolsep 43pt

\begin{tabular*}{\textwidth}{ccc}
\hline
reduction algorithm & lower bound \quad & upper bound  \\ \hline
{standard Kronecker substitution}  & {$d_{h_{x_n}}{\mathbb{D}}^{n-1}$} & {${\mathbb{D}}^n$}\\
iterative Kronecker substitution \quad  & $\prod_{\substack{i=1}}^nd_{h_{x_i}}$ & $ \prod_{\substack{i=1}}^n(d_{h_{x_i}}+1)$\\
CRT reduction &  $\max{ \left\{d_{h_{x_i}}\right\}}$ & $2M$\\
hybrid reduction & $\max{\left\{d_{h_{x_i}}\right\}}$ & $ \prod_{\substack{i=1}}^n(d_{h_{x_i}}+1)$\\
\hline
\end{tabular*}
\end{table}
{By Proposition~\ref{prop_sks}, the lower bound of $d_{h_x}^{\text{SKS}}$ is $d_{h_{x_n}}{\mathbb{D}}^{n-1}$, and the upper bound is ${\mathbb{D}}^n$.} By (\ref{eqn_ite_KS}), the lower bound of $d_{h_x}^{\text{IKS}}$ is $\prod_{\substack{i=1}}^nd_{h_{x_i}}$, and the upper bound is $ \prod_{\substack{i=1}}^n(d_{h_{x_i}}+1)$. The lower bound of $d_{h_x}^{\text{CRT}}$ is $\max{ \left\{ d_{h_{x_i}} \right\}}$, which can be achieved in some special cases where every monomial $x_1^{k_1}\cdots x_n^{k_n}$ satisfies $k_1=k_2=\cdots =k_n$, and its upper bound is $2M$ because every $n$-variable monomial contained in the original $f$ and $g$ is converted into the univariate monomial with every exponent of $x$ in $[0, M)$. Hybrid reduction combines the advantages of CRT reduction and  iterative Kronecker substitution because of the prediction of degree and the branch selection in Algorithm~\ref{alg_combine}.

Let $T_f$ and $T_g$ be the number of monomials contained in polynomials $f$ and $g$, respectively. Table~\ref{compare_compute} shows the computational complexity of the four reduction algorithms, \MS{where we focus on the computational complexity of the reduction from multivariate polynomials $f$ and $g$ to univariate polynomials.}
We regard the costs of multiplication, division and modulo to be equal and the costs of addition and subtraction to be equal.

\begin{table}[!htb]
\footnotesize
\caption{Computational complexity of the four reduction algorithms}
\label{compare_compute}
\tabcolsep 19pt
		\begin{tabular*}{\textwidth}{ccc}
			\hline
			reduction algorithm & number of multiplication operations  & number of addition operations \\
								      \hline	
			{standard Kronecker substitution}& {$\Theta(n(T_f+T_g))$} & {$\Theta(n(T_f+T_g))$} \\
			iterative Kronecker substitution& $\Theta(n(T_f+T_g))$ & $\Theta(n(T_f+T_g))$ \\
			CRT reduction & $O(2.078\ln{M}+2n(T_f+T_g))$ & $\Theta(n(T_f+T_g))$ \\
			hybrid reduction& $\Theta(n(T_f+T_g))$ & $\Theta(4n(T_f+T_g))$\\
			\hline
		\end{tabular*}
\end{table}

\MS{By (\ref{Eqn-SKS-Substition})}, 
the standard Kronecker substitution has $\Theta(n(T_f+T_g))$ multiplications and additions. The iterative Kronecker substitution has $\Theta(n(T_f+T_g))$ multiplications and additions according to Algorithm~\ref{alg_ite_KS}. \cite[Corollary $4.5.3L$]{21} indicates that calculating $\left(\frac{M}{p_i}\right)^{-1} \pmod{p_i}$ by Euclid's algorithm requires $O(2.078\ln{p_i})$ division steps; thus, there are
\[O\left(2.078\sum_{\substack{i=1}}^n{\ln{p_i}}\right)
=O\left(2.078\ln{M}\right)\]
divisions to obtain all $\left\{\left(\frac{M}{p_i}\right)^{-1}\right\}$. According to Algorithm~\ref{alg_crtsub}, there are $O(2.078\ln{M} +2n(T_f+T_g))$ multiplications and $\Theta(n(T_f+T_g))$ additions in the CRT reduction. In every iteration \MS{of} the hybrid reduction, the calculation of (\ref{eqn_dcrt}) requires $\Theta(T_f+T_g)$ subtractions; there are $\Theta(T_f+T_g)$ multiplications and $\Theta(3(T_f+T_g))$ additions in the CRT branch, \M{and there are} $\Theta(T_f+T_g)$ multiplications and $\Theta(T_f+T_g)$ additions in the iterative-Kronecker branch. Therefore, the hybrid reduction has $\Theta(n(T_f+T_g))$ multiplications and $\Theta(4n(T_f+T_g))$ additions.

\section{Conclusions}\label{sec_conclusions}
We adopt a new multivariate multiplication mechanism that involves the following: first, it reversibly reduces multivariate polynomials into univariate polynomials; then, it calculates the corresponding univariate product using a fast univariate multiplication method; and finally, it correctly recovers the multivariate product from the derived univariate product. Regarding the reversible reduction from multivariate polynomials to univariate polynomials, because the size of $d_{h_x}$ is the bottleneck of the subsequent univariate multiplication, we propose three reduction methods to minimize the obtained $d_{h_x}$. The first is iterative Kronecker substitution, which selects smaller substitution exponents compared with standard Kronecker substitution; the second is CRT reduction with a lower bound \W{of $d_{h_x}$}; and the last is hybrid reduction, combining the advantages of the other two methods. Experiments in the Appendix indicate that the proposed hybrid reduction can reduce $d_{h_x}$ even to approximately $3\%$ that obtained from standard Kronecker substitution on some randomly generated samples.

Admittedly, the contribution of the CRT branch is generally not as good as that of the iterative-Kronecker branch in the hybrid reduction but improves when $Max_f(i,j)$, $Max_f(j,i)$, $Max_g(i,j)$ and $Max_g(j,i)$ are small. Moreover, direct multiplication may be faster when polynomials are particularly sparse. Since different substitution sequences result in different $d_{h_x}$ values by iterative Kronecker substitution, the optimal substitution sequence is still under exploration.

\normalem  
\bibliographystyle{alpha}
\bibliography{refer}

\section*{Appendix}


We implement standard Kronecker substitution, iterative Kronecker substitution and hybrid reduction in two cases: {\textit{fully random case} and \textit{partially random case}.} {In both cases,} we randomly generate $4$-variable cases $f(x_1, x_2, x_3,x_4)$ and $g(x_1, x_2, x_3,x_4)$, with each containing 1,000,000 monomials (before merging similar monomials) and the tuple of degrees on ($x_1$, $x_2$, $x_3$, $x_4$) being (100, 100, 100, 100), (70, 80, 90 100), (40, 60, 80, 100) or (10, 40, 70, 100). However, we restrict the difference between the exponents of $x_1$ and $x_2$ to $[-L_{x_1,x_2},\ L_{x_1,x_2}]$ in every monomial in the partially random case, where $L_{x_1,x_2} \in N^{+}$. To compare the sizes of $d_{h_x}^{\text{IKS}}$ and $d_{h_x}^{\text{HR}}$ with $d_{h_x}^{\text{SKS}}$, we provide the ratios ${d_{h_x}^{\text{IKS}}}/{d_{h_x}^{\text{SKS}}}$ and ${d_{h_x}^{\text{HR}}}/{d_{h_x}^{\text{SKS}}}$. All values below are the averages of twenty experiments.

\begin{table}[!htb]
\footnotesize
\caption{Ratios ${d_{h_x}^{\text{IKS}}}/{d_{h_x}^{\text{SKS}}}$ and ${d_{h_x}^{\text{HR}}}/{d_{h_x}^{\text{SKS}}}$ at different tuples of degrees, in the fully random case}
\label{speedup_case1}
\tabcolsep 16pt
		\begin{tabular*}{\textwidth}{ccccc}
			\hline
			tuple of degrees & (100, 100, 100, 100) & (70, 80, 90, 100) & (40, 60, 80, 100) & (10, 40, 70, 100) \\ \hline
			${d_{h_x}^{\text{IKS}}}/{d_{h_x}^{\text{SKS}}}$ & 1.000 & 0.506  & 0.195  & 0.030 \\
			${d_{h_x}^{\text{HR}}}/{d_{h_x}^{\text{SKS}}}$ & 1.000 & 0.506  & 0.195  & 0.030 \\
			\hline
		\end{tabular*}
\end{table}
Table~\ref{speedup_case1} shows the ratios ${d_{h_x}^{\text{IKS}}}/{d_{h_x}^{\text{SKS}}}$ and ${d_{h_x}^{\text{HR}}}/{d_{h_x}^{\text{SKS}}}$ in the fully random case, regarding which we give an intuitive explanation as follows.

In this case, for all $x_i \neq x_j$, we have
\[Pr\left(\exists \left(x_i^{a}x_j^{b}\cdots \right)  \in  f(x_1,\dots ,x_4) \right) =1-\left(1-\frac{1}{\left(d_{f_{x_i}}+1\right)\left(d_{f_{x_j}}+1\right)}\right)^{1000000} \ \textrm{for some constant} \ a \ \textrm{and} \ b. \] 
We substitute $x_2,x_3,x_4$ with $x_1$ in turns; thus, in the \MS{$(j-1)$-th} iteration of the hybrid reduction, we have $x_i=x_1$, and
\[\deg_{x_1}f(x_1,x_1^{D_2},\dots , x_1^{D_{j-1}},x_j,\dots ,x_4)= \max\big\{k_1+k_2 \times D_2+\cdots +k_{j-1} \times D_{j-1} \ | \
x_1^{k_1} \cdots x_{j-1}^{k_{j-1}}\cdots \in f(x_1, \dots, x_4)\big\},\]
\[\! \mathit{Max}_{f(x_1,x_1^{D_2},\dots , x_1^{D_{j-1}},x_j,\dots ,x_4  )}(j,1) \!=\!\max \big\{k_1+k_2 \times D_2+\cdots +k_{j-1} \times D_{j-1} - k_j \ | \
x_1^{k_1} \cdots x_{j-1}^{k_{j-1}}x_j^{k_j}\cdots \in f(x_1, \dots, x_4)\big\};\]
moreover, with high probability $\exists \left(x_{j-1}^{d_{f_{x_{j-1}}}}x_j^{0}\cdots \right)  \in  f(x_1,\dots ,x_4) $, and
\[D_{j-1}> \max\big\{k_1+k_2 \times D_2+\cdots +k_{j-2} \times D_{j-2} \ | \
x_1^{k_1} \cdots x_{j-2}^{k_{j-2}}\cdots \in f(x_1, \dots, x_4)\big\},\]
$\max \Big\{k_{j-1} \times D_{j-1} \ | \ x_{j-1}^{k_{j-1}}\cdots \in f(x_1, \dots, x_4)\Big\} \approx  d_{f_{x_{j-1}}} \times D_{j-1}$ is the dominant \MS{factor} in both $\deg_{x_1}f(x_1, \\ x_1^{D_2},\dots , x_1^{D_{j-1}},x_j,\dots ,x_4)$ and $\mathit{Max}_{f(x_1,x_1^{D_2},\dots , x_1^{D_{j-1}},x_j,\dots ,x_4  )}(j,1)$. \MS{Therefore,}
\begin{eqnarray}\label{eqn_max_i1}
\frac{\mathit{Max}_{f(x_1,x_1^{D_2},\dots , x_1^{D_{j-1}},x_j,\dots ,x_4  )}(j,1)}{\deg_{x_1}f(x_1,x_1^{D_2},\dots , x_1^{D_{j-1}},x_j,\dots ,x_4)}\approx 1.
\end{eqnarray}
In addition, 
\begin{equation}\label{eqn_max_1i_upper}
\mathit{Max}_{f(x_1,x_1^{D_2},\dots , x_1^{D_{j-1}},x_j,\dots ,x_4  )}(1,j) \leq \deg_{x_j}f(x_1,x_1^{D_2},\dots , x_1^{D_{j-1}},x_j,\dots ,x_4),
\end{equation}
\begin{eqnarray}\label{eqn_max_1i}
\mathit{Max}_{f(x_1,x_1^{D_2},\dots , x_1^{D_{j-1}},x_j,\dots ,x_4  )}(1,j) \!\!\!\! &=& \!\!\!\! \max \Big\{k_j - (k_1+k_2 \times D_2+\cdots +k_{j-1} \times D_{j-1}) \ | \ x_1^{k_1} \cdots x_{j-1}^{k_{j-1}}x_j^{k_j}\cdots  \nonumber \\ \in f(x_1, \dots, x_4)\Big\}
\!\!\!\!&\geq&\!\!\!\! -\min \big\{ k_1+\cdots +k_{j-1} \times D_{j-1} \ | \
x_1^{k_1} \cdots x_{j-1}^{k_{j-1}}\cdots \in f(x_1, \dots, x_4)\big\}.
\end{eqnarray}
If \W{$\min\{k_{j-1} \ | \
x_{j-1}^{k_{j-1}}\cdots \in f(x_1, \dots, x_4) \}$} is small (with high probability, $\exists \left(x_{j-1}^{0}\cdots \right)  \in  f(x_1,\dots ,x_4)$), with high probability, $\min \big\{ k_1+\cdots +k_{j-1} \times D_{j-1} \ | \
x_1^{k_1} \cdots x_{j-1}^{k_{j-1}}\cdots \in f(x_1, \dots, x_4)\big\}$ is much smaller than $\mathit{Max}_{f(x_1,x_1^{D_2},\dots , x_1^{D_{j-1}},x_j,\dots ,x_4  )}(j,1)$. Thus, from (\ref{eqn_max_1i_upper}) and (\ref{eqn_max_1i}), either $\mathit{Max}_{f(x_1,x_1^{D_2},\dots , x_1^{D_{j-1}},x_j,\dots ,x_4  )}(1,j)>0$ or $\big|\mathit{Max}_{f(x_1,x_1^{D_2},\dots , x_1^{D_{j-1}},x_j,\dots ,x_4  )}(1,j)\big|$ is much smaller than $\mathit{Max}_{f(x_1,x_1^{D_2},\dots , x_1^{D_{j-1}},x_j,\dots ,x_4  )}(j,1)$,
and similarly for $g$. From (\ref{eqn_max_i1}), (\ref{eqn_max_1i_upper}) and (\ref{eqn_max_1i}), for our case, with $x_i=x_1$, (\ref{eqn_dcrt}) is greater than $\deg_{x_1}f(x_1,x_1^{D_2},\dots , x_1^{D_{j-1}},x_j,\dots ,x_4) \times \deg_{x_j}f(x_1,x_1^{D_2},  \dots , x_1^{D_{j-1}}, \\ x_j,\dots ,x_4)$.
Therefore, hybrid reduction selects the iterative-Kronecker branch in every iteration; 
then, ${d_{h_x}^{\textrm{HR}}}$ is equal to ${d_{h_x}^{\textrm{IKS}}}$, \ms{which explains the two identical lines in Table \ref{speedup_case1},}
and
\begin{equation}\label{eqn_ratios}
\frac{d_{h_x}^{\textrm{IKS}}}{d_{h_x}^{\textrm{SKS}}}= \frac{d_{h_x}^{\textrm{HR}}}{d_{h_x}^{\textrm{SKS}}} \approx  \frac{\prod_{i=1}^n \left(d_{h_{x_i}}+1\right)}{d_{h_{x_n}}{\mathbb{D}}^{n-1}},
\end{equation}
e.g., the ratios are
\[\frac{21\times 81\times 141\times 201}{200\times 201^3} \approx  0.030\]
when the tuple of degrees is (10, 40, 70, 100).

\begin{figure}[!htb]
\footnotesize
\centering
\begin{minipage}[c]{0.25\textwidth}
\centering
\includegraphics[trim={0cm 0cm 0cm 0cm},clip,width=\textwidth]{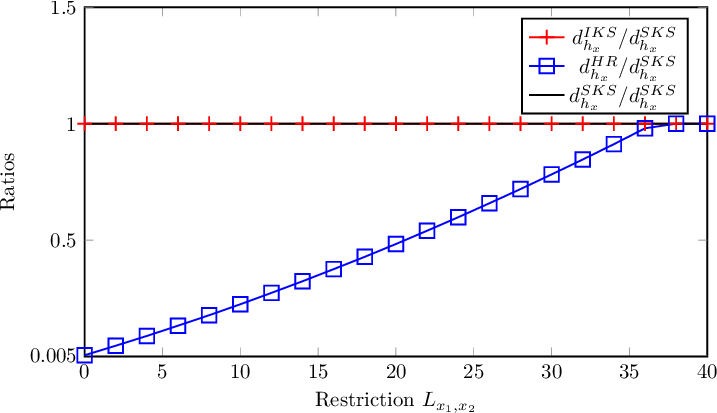} \\ {\quad \quad (a) The tuple of degrees is (100, 100, 100, 100)} \\ \quad
\label{fig_100}
\end{minipage}
\hspace{0.02\textwidth}
\begin{minipage}[c]{0.25\textwidth}
\centering
\includegraphics[trim={0cm 0cm 0cm 0cm},clip,width=\textwidth]{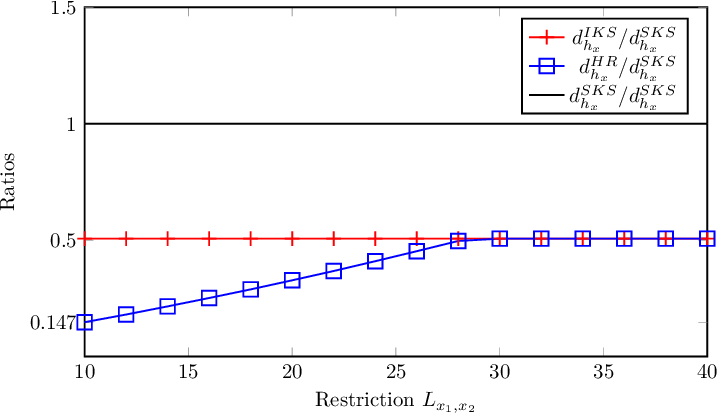} \\ {\quad \quad (b) The tuple of degrees is (70, 80, 90, 100)} \\ \quad
\label{fig_70}
\end{minipage}

\begin{minipage}[c]{0.25\textwidth}
\centering
\includegraphics[trim={0cm 0cm 0cm 0cm},clip,width=\textwidth]{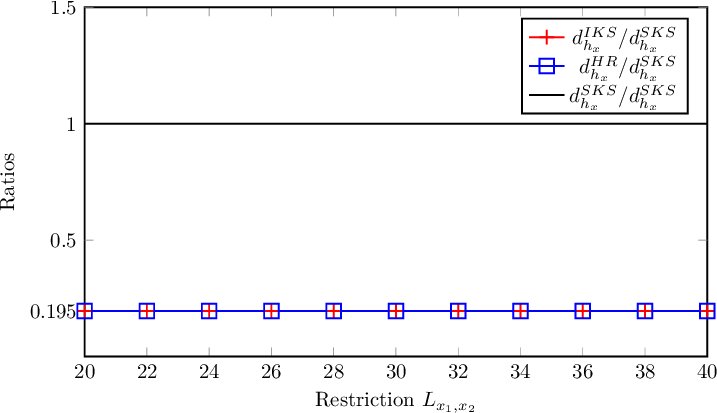} \\{\quad \quad (c) The tuple of degrees is (40, 60, 80, 100)}\\ \quad
\label{fig_40}
\end{minipage}
\hspace{0.02\textwidth}
\begin{minipage}[c]{0.25\textwidth}
\centering
\includegraphics[trim={0cm 0cm 0cm 0cm},clip,width=\textwidth]{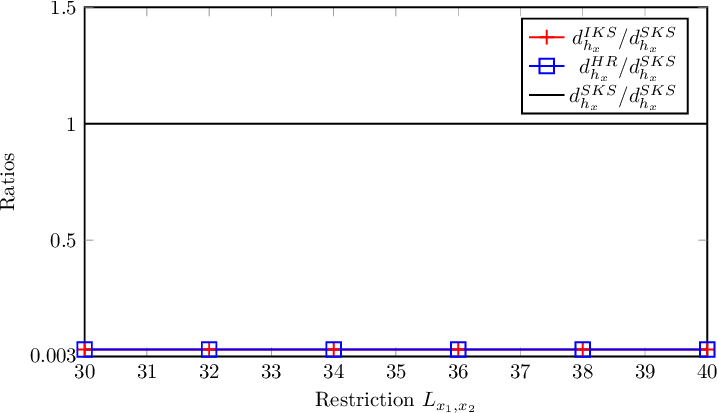} \\ {\quad \quad (d) The tuple of degrees is (10, 40, 70, 100)}\\ \quad
\label{fig_10}
\end{minipage}
\caption{Ratios ${d_{h_x}^{\textrm{IKS}}}/{d_{h_x}^{\textrm{SKS}}}$ and ${d_{h_x}^{\textrm{HR}}}/{d_{h_x}^{\textrm{SKS}}}$ at different tuples of degrees in the partially random case}
\label{fig_ratio_2}
\end{figure}

{
Figure~\ref{fig_ratio_2} shows the ratios of ${d_{h_x}^{\text{IKS}}}/{d_{h_x}^{\text{SKS}}}$ and ${d_{h_x}^{\text{HR}}}/{d_{h_x}^{\text{SKS}}}$ in the partially random case, regarding which we provide an intuitive explanation in the following. In this case, the ratio ${d_{h_x}^{\text{IKS}}}/{d_{h_x}^{\text{SKS}}}$ remains the same and is equal to (\ref{eqn_ratios}) as $L_{x_1,x_2}$ increases because substitution exponents $\left\{D_i\right\}$ from (\ref{eqn_iksd}) and $\left\{\mathbb{D}_i=\mathbb{D}^{i-1}\right\}$ from (\ref{eqn_D}) are hardly affected by the size of $L_{x_1,x_2}$.
In this case,
$\mathit{Max}_f(1,2)=\mathit{Max}_f(2,1)=\mathit{Max}_g(1,2) =\mathit{Max}_g(2,1)=L_{x_1,x_2}$ because of the restriction in the partially random case. Therefore, if $L_{x_1,x_2}$ is relatively small such that (\ref{eqn_dcrt}) is smaller \W{than $d_{h_{x_1}} \times d_{h_{x_2}}$}, hybrid reduction will select the CRT branch to reduce $x_1$ and $x_2$, resulting in
\begin{equation}\label{eqn_ratios2}
\frac{d_{h_x}^{\text{HR}}}{d_{h_x}^{\text{SKS}}} \approx  \frac{\left(4L_{x_1,x_2}(d_{h_{x_2}}+2L_{x_1,x_2}+2) +1\right)\prod_{i=3}^n \left(d_{h_{x_i}}+1\right)}{d_{h_{x_n}}{\mathbb{D}}^{n-1}}.
\end{equation}
Otherwise, the ratio ${d_{h_x}^{\text{HR}}}/{d_{h_x}^{\text{SKS}}}$ is equal to (\ref{eqn_ratios}). In Figure~\ref{fig_ratio_2}(a) and~\ref{fig_ratio_2}(b), the ratio ${d_{h_x}^{\text{HR}}}/{d_{h_x}^{\text{SKS}}}$ increases first, with the value being (\ref{eqn_ratios2}), and then remains the same, with the value being  (\ref{eqn_ratios}). In Figure~\ref{fig_ratio_2}(c) and~\ref{fig_ratio_2}(d), $L_{x_1,x_2}$ must be no less than 20 in Figure~\ref{fig_ratio_2}(c) and no less than 30 in Figure~\ref{fig_ratio_2}(d), i.e., $L_{x_1,x_2}$ is large such that hybrid reduction always selects the iterative-Kronecker branch. Then, the ratio remains the same and is equal to (\ref{eqn_ratios}) as $L_{x_1,x_2}$ increases.
}

From Table~\ref{speedup_case1} and Figure~\ref{fig_ratio_2}, hybrid reduction is more effective than standard Kronecker substitution, and the CRT branch is very effective when $L_{x_1,x_2}$ is relatively small. For polynomials generated in the fully random case and having many monomials, the contribution of the CRT branch is small, and the efficiency of hybrid reduction mainly comes from the iterative-Kronecker branch.

%
%
%
%
%
%
%
%

\end{document}